\documentclass[fleqn,usenatbib,onecolumn]{mnras}

\usepackage{newtxtext,newtxmath}

\usepackage[T1]{fontenc}

\DeclareRobustCommand{\VAN}[3]{#2}
\let\VANthebibliography\thebibliography
\def\thebibliography{\DeclareRobustCommand{\VAN}[3]{##3}\VANthebibliography}

\usepackage{graphicx}	
\usepackage{amsmath}	
\usepackage{orcidlink}	

\newcommand{\rmd}{{\rm d}}
\newcommand{\rme}{{\rm e}}
\newcommand{\rmi}{{\rm i}}

\newcommand{\mbfr}{\mathbf{r}}

\title[Scattering model of scintillation arcs]{Scattering model of scintillation arcs in pulsar secondary spectra}

\author[T. Kramer et al.]{
Tobias Kramer,$^{1,2}$\orcidlink{0000-0003-1106-3587}
Daniel Waltner,$^{1}$\orcidlink{0000-0002-3036-6463}
Eric J.\ Heller,$^{2}$\orcidlink{0000-0002-5398-0861}
and Dan R.\ Stinebring$^{3}$\orcidlink{0000-0002-1797-3277}
\\
$^{1}$Institute for Theoretical Physics, Johannes Kepler University Linz, Altenberger Str.\ 69, 4040 Linz, Austria\\
$^{2}$Harvard Physics Department, 17 Oxford Street, Cambridge, MA 02138, USA\\
$^{3}$Department of Physics and Astronomy, 110 No. Professor St., Oberlin College, Oberlin, OH 44074, USA
}

\pubyear{2024}

\begin{document}
\label{firstpage}
\maketitle

\begin{abstract}
The dynamic spectra of pulsars frequently exhibit diverse interference patterns, often associated with parabolic arcs in the Fourier-transformed (secondary) spectra.
Our approach differs from previous ones in two ways:  
first, we extend beyond the traditional Fresnel-Kirchhoff method by using the Green's function of the Helmholtz equation, i.e.\ we consider spherical waves originating from three dimensional space, not from a two dimensional screen.
Secondly, the discrete structures observed in the secondary spectrum result from discrete scatterer configurations, namely plasma concentrations in the interstellar medium, and not from the selection of points by the stationary phase approximation.
Through advanced numerical techniques, we model both the dynamic and secondary spectra, providing a comprehensive framework that describes all components of the latter spectra in terms of physical quantities.
Additionally, we provide a thorough analytical explanation of the secondary spectrum.
\end{abstract}




\section{Introduction} \label{sec:intro}

The first observation of radio pulsars goes back to \cite{hewish_observation_1969}.
The electromagnetic signals of pulsars encounter on their way a varying electron density of the interstellar medium (ISM), resulting in deflection and scattering of the travelling electromagnetic waves. 
In addition, the relative motion of pulsar, ISM, and the observer's antenna leads to a Doppler shift and time-varying phenomena.
The dynamic spectra consist of the frequency resolved pulse sequences observed over a time-span of up to several hours.
A two-dimensional (2D) Fourier transform of the dynamic spectra gives the secondary spectra.
\cite{Stinebring2001} discovered parabolic arc structures in the secondary spectra, which result from the interference of multiple signal pathways at a specific distance between the pulsar and the observer.
The recent catalogue of scintillation arcs compiled by \cite{Stinebring2022} of 22 pulsars shows various structures and contains the basic physical parameters of the pulsar, such as distance and velocity.
\cite{Walker2004} developed theoretical descriptions of the parabolic arcs starting from the Fresnel-Kirchhoff integral.
Using Monte Carlo methods, \cite{Walker2004} then computed the locations of points in the secondary spectra.
Similarly, \cite{cordes_theory_2006} used a thin phase-changing screen approach to study the dynamic and secondary spectra. Since then there have been a variety of arc studies based on observations by different groups, e.g., \cite{hill_pulsar_2003,hill_deflection_2005,wang_long-term_2005,bhat_scintillation_2016,safutdinov_secondary_2017,wang_jiamusi_2018,Stinebring2019,reardon_precision_2020,rickett_scintillation_2021,yao_evidence_2021,chen_interstellar_2022,mckee_probing_2022}.

Here we put forward a different theoretical approach to treat scattering by the ISM using Green's functions.
This method is commonly applied to scattering problems in quantum mechanics; see \cite{Kramer2006} for an application to matter waves originating from a compact source.
By solving Helmholtz's equation in Cartesian coordinates using Green's functions we determine the pulsar spectra received after scattering at the interstellar medium (dynamic spectrum) and its 2D Fourier transform with respect to time and frequency by high-precision numerics (secondary spectrum) for a given scattering configuration. 
In contrast to the Fresnel-Kirchhoff approach, our method enables the determination of  the entire spectrum and relates the strengths of the individual components to physical quantities such as the refractive index and the wavenumber. 
Furthermore, we give a complete analytical description of the secondary spectra. 
\cite{Walker2004} obtained point-like peaks in the snapshot regime and determined their positions. We considerably extend this analysis by analytically determining also the peak extensions and the intensities.

In the second section we introduce our scattering approach and present our findings for an analytical description of the spectra in the third section. 
We conclude in the fourth section and relegate to the appendices some detailed explanations and technical details. 

\section{Solution of the Helmholtz equation}

\begin{figure}
    \centering
    \includegraphics[width=0.99\textwidth]{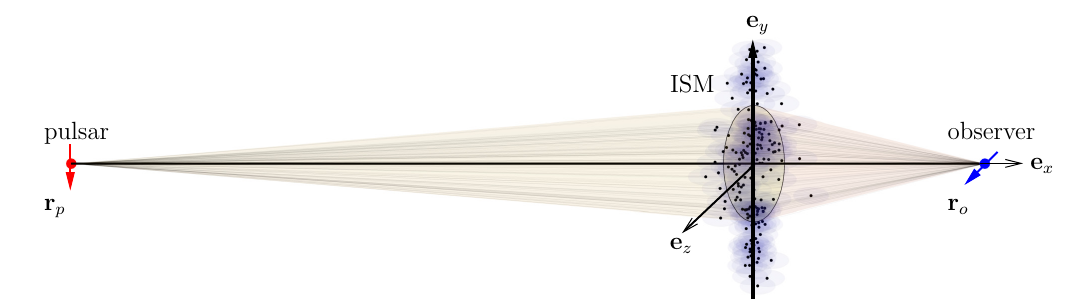}
    \caption{Sketch of the scattering setup, including the pulsar, the ISM, and the observer on Earth.
    The drawing is not to scale, the extension of the ISM along the vertical axis is about $10^9$ times exaggerated.
    We take the ISM to be at rest, while the pulsar and the observer are possibly moving in orthogonal directions with respect to each other and the connecting line observer -- pulsar.
    The coordinate origin is taken to be at the intersection of the ISM and the line of sight.
    The cones mark the scattering disk.
    Within the Born approximation structures within the ISM are contracted to point scatterers (black dots).
    }
    \label{fig:pulsarscheme}
\end{figure}

In this section we propose a Green's function method to describe the scattering of pulsar radiation in the ISM. 
The differences to the Fresnel-Kirchhoff approach are summarised in Appendix~\ref{app1}. 
We consider scattering from an extended plasma cloud (see Fig.~\ref{fig:pulsarscheme}), described by a region with the scattering potential $V(\mathbf{r}')$
\begin{equation}
V(\mathbf{r}')=\frac{1}{4\pi}(\epsilon(\mathbf{r}')-\epsilon_{\rm background}) \ne 0.
\end{equation}
The electron number density $n_e$ of the plasma cloud determines the plasma frequency $\omega_p$, the refractive index $n$, and $\epsilon$:
\begin{equation}
    \epsilon(\mathbf{r'}) = n^2(\mathbf{r'}) = 1-\frac{\omega_p^2}{\omega^2}, \quad \omega_p^2=\frac{n_e(\mathbf{r'}) e^2}{\epsilon_0 m_{e}}.
\end{equation}
Pulsar signals travel long distances and undergo a dispersion due to the average electron density in the galaxy.
This results in pulses where different frequency components arrive at different times.
Here, we are interested in the ISM properties affecting the signal on shorter scales compared to the pulsar distance.
This allows us to set the dielectric constant of the background to unity, but alternatively, a uniform background could be introduced.
Maxwell's equation for the electric field becomes \cite[Eq.~(50.35)]{schwinger_classical_1998}
\begin{equation}\label{eq:efield}
    \mathbf{E}(\mathbf{r})=\mathbf{E}_\text{inc}(\mathbf{r})+\rmi k (\mathbf{1}+\frac{1}{k^2}\nabla\nabla^T)\cdot \int \rmd\mathbf{r}' G_0(\mathbf{r},\mathbf{r}';k)(-\rmi k)  V(\mathbf{r'}) \mathbf{E}(\mathbf{r}').
\end{equation}
The free Green's function reads
\begin{equation}
G_0(\mathbf{r},\mathbf{r}';k)=\frac{\rme^{i k |\mathbf{r}-\mathbf{r}'|}}{|\mathbf{r}-\mathbf{r}'|}, \quad k=\frac{\omega}{c}=\frac{2\pi\nu}{c}.
\end{equation}
Within the Born approximation we replace in the integral the electric field by the incoming electric field $\mathbf{E}_\text{inc}(\mathbf{r}')$ and also neglect any change in the polarisation direction by dropping the Hessian in the second term in Eq.~(\ref{eq:efield}). 
This can be easily verified considering e.g.\ a linearly polarised wave possessing solely a non-vanishing $y$ component depending only on the spatial $x$ coordinate.
We take the incoming electric field to be a spherical wave emitted from the pulsar
\begin{equation}
    \mathbf{E}_\text{inc}(\mathbf{r})=\mathbf{U}_0 G_0(\mathbf{r},\mathbf{r}_p;k),
\end{equation}
where $\mathbf{U}_0$ determines the polarisation direction of the electric field and has units of a voltage.
We obtain the electric field from the Green's function
\begin{equation}\label{eq:greenE}
    \mathbf{E}(\mathbf{r})=\mathbf{U}_0 G_0(\mathbf{r},\mathbf{r}_p;k)+k^2 \int \rmd\mathbf{r}' G_0(\mathbf{r},\mathbf{r}';k)\; V(\mathbf{r'}) \; \mathbf{U}_0 G_0(\mathbf{r}',\mathbf{r}_p;k),
\end{equation}
where we consider only one interaction with the scattering potential $V$, corresponding to the Born approximation.
The electric field in the Born approximation consists of two contributions, first the unscattered component in the absence of any medium, and second the volume integral over the distribution of plasma clouds in the interstellar medium.
For dense or compact objects multiple scattering could be included by summing the Born series in terms of the transition matrix.
Such processes are neglected in Eq.~(\ref{eq:greenE}).
This equation is also used in quantum mechanics to describe the scattering of coherent electron waves at obstacles \cite[ch.~26]{heller_semiclassical_2018}.
Since we are only interested in the relative contributions of the electromagnetic waves we set $|\mathbf{U}_0|=1$ with direction orthogonal to the direction of propagation.
The intensity is given by the absolute value squared of the electric field
\begin{equation}
 H(\mbfr,\mbfr_p;\nu)={\left|E(\mbfr)\right|}^2=
 {\left|
G_0(\mathbf{r},\mathbf{r}_p;k)+k^2 \int \rmd\mathbf{r}' G_0(\mathbf{r},\mathbf{r}';k)\; V(\mathbf{r'}) \; G_0(\mathbf{r}',\mathbf{r}_p;k)
\right|}^2.\label{eq:H}
\end{equation}
By taking the absolute value, interference terms appear in the exponents related to the free Green's function.
The argument of the exponent contains the differences in distances measured in multiples of the wavelength.
For the typical pulsar geometry shown in Fig.~\ref{fig:pulsarscheme} all path differences are slowly varying functions across the interstellar medium.
We introduce coarse-grained integration regions, which result in a collection of three-dimensional clouds of scattering sources across an entire region of the ISM.
For simplicity, we consider a Gaussian electron density profile of the $i$th cloud centred around $\mathbf{r}_i$ of the form
\begin{equation}
    n_{e,i}(\mathbf{r})=n_{e,\text{peak},i} \exp\left(-\frac{{(\mathbf{r}-\mathbf{r}_i)}^2}{2 a^2}\right),
\end{equation}
where $n_{e,\text{peak}}$ denotes the peak electron density.
The first order Born approximation requires to evaluate
\begin{align}
    H(\mbfr_o,\mbfr_p;\nu)&=\label{eq:H1}
    {
    \left|
    G_0(\mathbf{r}_o,\mathbf{r}_p;k)+k^2
    \sum_{i}
    \int_{\text{cloud}_i} \rmd\mathbf{r}' 
    \frac{\rme^{\rmi k |\mathbf{r}_o-\mathbf{r}'|}}{|\mathbf{r}_o-\mathbf{r}'|}V(\mathbf{r}')
    \frac{\rme^{\rmi k |\mathbf{r}_p-\mathbf{r}'|}}{|\mathbf{r}_p-\mathbf{r}'|}\right|}^2\\\notag
    &\approx
    \frac{1}{{|\mathbf{r}_p-\mathbf{r}_o|}^2}
    -
    \sum_{i}
    \beta_i
    \frac{\rme^{ \rmi k ( |\mathbf{r}_p-\mathbf{r}_i|
                     +|\mathbf{r}_i-\mathbf{r}_o|
                     -|\mathbf{r}_p-\mathbf{r}_o|)
                     }+\rme^{ -\rmi k ( |\mathbf{r}_p-\mathbf{r}_i|
                     +|\mathbf{r}_i-\mathbf{r}_o|
                     -|\mathbf{r}_p-\mathbf{r}_o|)
                     }}
                     {|\mathbf{r}_p-\mathbf{r}_i| |\mathbf{r}_i-\mathbf{r}_o| |\mathbf{r}_p-\mathbf{r}_o|} \nonumber\\
    &+
    \sum_{i,j}
   \beta_i\beta_j
    \frac{\rme^{ \rmi k ( |\mathbf{r}_p-\mathbf{r}_i|
                     +|\mathbf{r}_i-\mathbf{r}_o|
                     -|\mathbf{r}_p-\mathbf{r}_j|
                     -|\mathbf{r}_j-\mathbf{r}_o| ) }}
                     {|\mathbf{r}_p-\mathbf{r}_i| |\mathbf{r}_i-\mathbf{r}_o| |\mathbf{r}_p-\mathbf{r}_j| |\mathbf{r}_j-\mathbf{r}_o|}.\label{eq:pspec}
\end{align}
In the last step in the equation above we evaluated the Gaussian integral via a series expansion for $a$ (see appendix~\ref{app:Born}) and introduced the parameter $\beta_i$
\begin{equation}\label{eq:eta}
    \beta_i =-k^2 \int_{\text{cloud}_i} \!\!\!\rmd\mathbf{r}' V(\mathbf{r}')
    =\frac{1}{4\pi}\frac{\omega^2}{c^2} \int_{\text{cloud}_i} \!\!\!\rmd\mathbf{r}'  \frac{\omega_p^2(\mathbf{r}')}{\omega^2}
    =\frac{e^2}{4\pi\epsilon_0 m_{e} c^2} \int_{\text{cloud}_i} \!\!\!\rmd\mathbf{r}'  n_{e,i}(\mathbf{r}')
    =\sqrt{\frac{\pi}{2}} \frac{a^3 e^2}{\epsilon_0 m_e c^2} n_{e,\text{peak},i}.
\end{equation}
A uniform background density $n_{e,b}$ along the signal propagation requires to replace the electron density with the local change in density $n_e(\mathbf{r}') \rightarrow (n_e(\mathbf{r}')-n_{e,b})$.
Within the Fresnel-Kirchhoff approach in \cite{Walker2004}, all the contributions in (\ref{eq:H}) and (\ref{eq:pspec}) are treated on equal footing rendering it impossible to distinguish their individual prefactors. 

The coarse-graining of the ISM complements other approaches which focus on the correlation function of scattering from extended sources described in terms of statistical spatial correlation functions, see \cite{tatarski_wave_1961,coles_scattering_2010}.

\begin{figure}
\begin{center}
\includegraphics[width=0.33\textwidth]{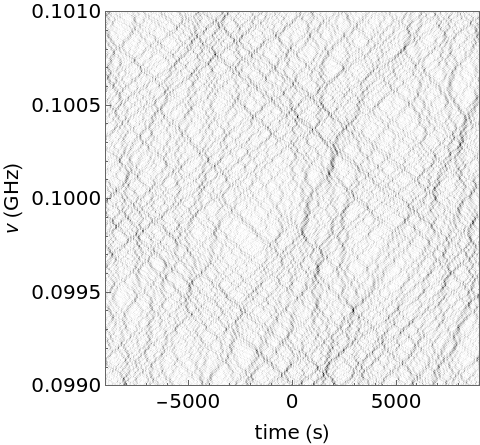}
\includegraphics[width=0.4\textwidth]{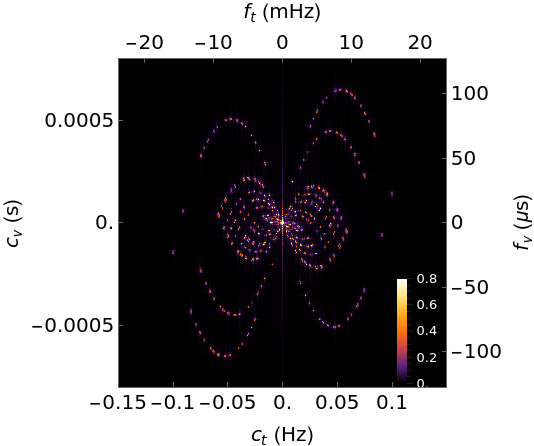}\\
\includegraphics[width=0.37\textwidth]{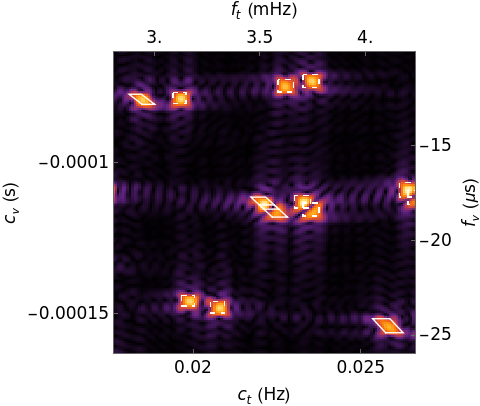}
\includegraphics[width=0.37\textwidth]{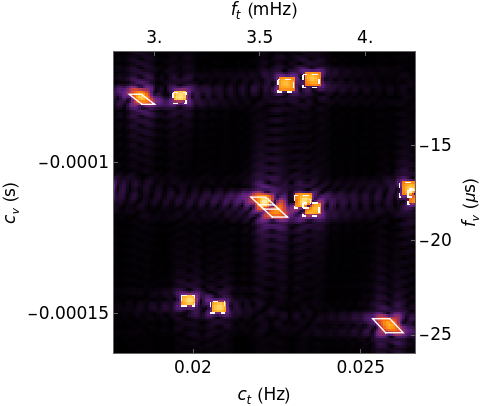}
  \end{center}
  \caption{Theoretical dynamic spectra $H(t,\nu)$, upper left panel, and secondary spectra ($|\tilde{H}(c_t=2\pi f_t,c_\nu=2\pi f_\nu)|$, upper right panel.
  The lower row panels show the secondary spectra in detail (left panel: numerical evaluation of Eq.~(\ref{eq:secspec}), right panel: analytic result from Eqs.~(\ref{eq:secspecanalyticmain}), (\ref{eq:ei}).
  The polygons indicate the analytical boundaries of the main arc features (solid lines) and inverted arcs (dashed lines).
  The centre frequency is set at $\nu_c=0.1$~GHz to produce large features in the secondary spectra.
  All scatterers are located on a line perpendicular to the line of sight, intersecting the line of sight.
  ($\beta=4.5\times 10^{18}$~m, $x_p=-214$~pc, $x_o=429$~pc, $v_p=640$~km/s).
  }\label{fig:spectra}
\end{figure}

\section{Dynamic and secondary spectra}
\subsection{Dynamic spectra}
Eq.~(\ref{eq:pspec}) contains the complete description of the electric field and is next evaluated for specific conditions.
We consider a coordinate system where the ISM is considered to be at rest and distributed around the origin of the coordinate system, while the pulsar and the observer are moving as shown in Fig. \ref{fig:pulsarscheme}.
The pulsar is moving with velocity $\mathbf{v}_p$ and changes position as function of time
\begin{equation}
\mathbf{r}_p(t)=\mathbf{r}_p(0)+\mathbf{v}_p t,
\end{equation}
likewise the observer moves with velocity $\mathbf{v}_o$ and is located at  position
\begin{equation}
\mathbf{r}_o(t)=\mathbf{r}_o(0)+\mathbf{v}_o t.
\end{equation}
The line of sight vector lies along $\mathbf{e}_x$ and is given by the direct connection of $\mathbf{r}_o(0)$ and $\mathbf{r}_p(0)$, the effect of the velocity components along this axis is negligible due to large values of the distance $|x_p|$ between the pulsar and the ISM plane and the distance $|x_o|$ between the ISM plane and the observer. 
Therefore the velocity component of $\mathbf{v}_p$  along $\mathbf{e}_y$ is the only relevant one, while for $\mathbf{v}_o$ the components along $\mathbf{e}_y$ and $\mathbf{e}_z$ need to be considered. 
A dynamic spectrum is obtained by recording $H(\mbfr_o(0)+\mathbf{v}_ot,\mbfr_p(0)+\mathbf{v}_p t;\nu)$ over the time domain $[-\Delta_t/2,\Delta_t/2]$ and frequency domain $[\nu_c-\Delta_\nu/2,\nu_c+\Delta_\nu/2]$. 
Computed spectra are shown in  Fig.~\ref{fig:spectra}, left panel, obtained from numerically evaluating Eq.~(\ref{eq:H}) and the analytical formulae for 25 scattering clouds.
Each scattering cloud is assigned the same value of $\beta_i=4.5\times 10^{18}$~m $(i=1,\ldots,25)$.
One possible set of parameters for the Gaussian cloud model is $a=10^9$~m and $n_{e,\text{peak}}=0.1$~cm$^{-3}$.
Further parameters are given in the caption of Fig.~\ref{fig:spectra}.
We used the extended precision mathematical functions available in Mathematica, \cite{wolfram_research_inc_mathematica_2024}, and the GCC Quad-Precision Math Library, \cite{free_software_foundation_gcc_2024} as we needed to determine trigonometric functions of large arguments.

\subsection{Secondary spectra}

The secondary spectrum is given by the two-dimensional Fourier transform of the preceding expression
\begin{equation}\label{eq:secspec}
  \left| \tilde{H}(c_t,c_\nu) \right|=
  \left| {\cal F}\left[H(\mbfr_o(t),\mbfr_p(t);\nu)\right] \right| 
  \propto 
\left|
 \int_{{\nu_c}-\Delta_\nu/2}^{{\nu_c}+\Delta_\nu/2}\int_{-\Delta_t/2}^{\Delta_t/2} \rme^{\rmi \nu c_\nu+\rmi t c_t}
  H(\mbfr_o(0)+\mathbf{v}_ot,\mbfr_p(0)+\mathbf{v}_p t;\nu) \; \rmd\nu \; \rmd t 
  \right|, 
\end{equation}
We note that the secondary spectrum is conventionally defined as the square of the latter quantity. 
Our usage here coincides with the ``conjugate spectrum” used by other authors (e.g.\ \cite{simard_disentangling_2019}), although we are considering only the magnitude of that quantity.
Often this quantity is also shown on a logarithmic scale (see  Fig.~\ref{fig:logscale}), in that case $\left| \tilde{H}(c_t,c_\nu) \right|$ and $\left| \tilde{H}(c_t,c_\nu) \right|^2$ differ only by a factor 2.
The secondary spectrum is shown in Fig.~\ref{fig:spectra}.
It is obtained by the discrete Fourier transform of $(512,512)$ points of the dynamic spectra which are zero padded to size $(1536,1536)$.
The secondary spectra show a wealth of sharply delineated features, caused by the presence of the ISM and the specific Fourier integration domain.
All results shown in the figures result from a numerical evaluation of Eq.~(\ref{eq:pspec}) and Eq.~(\ref{eq:secspec}).
For the interpretation of the numerical results, we discuss different levels of approximations of the integrals in the following sections.

\begin{figure}
\begin{center}
\includegraphics[width=0.33\textwidth]{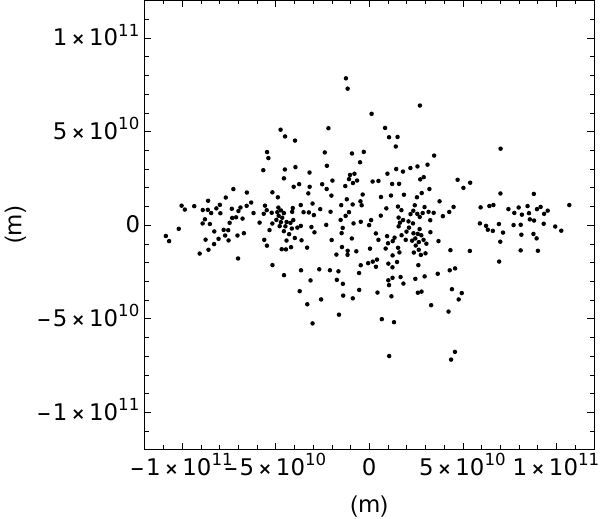}
\includegraphics[width=0.31\textwidth]{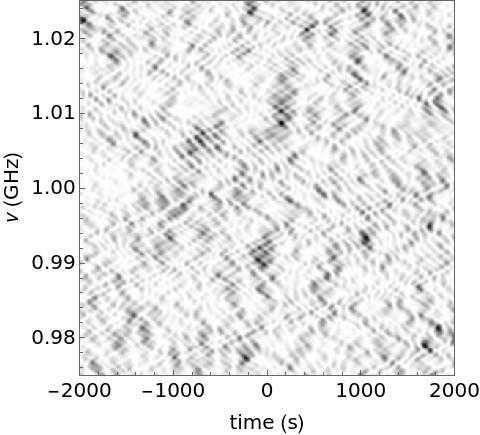}
\includegraphics[width=0.351\textwidth]{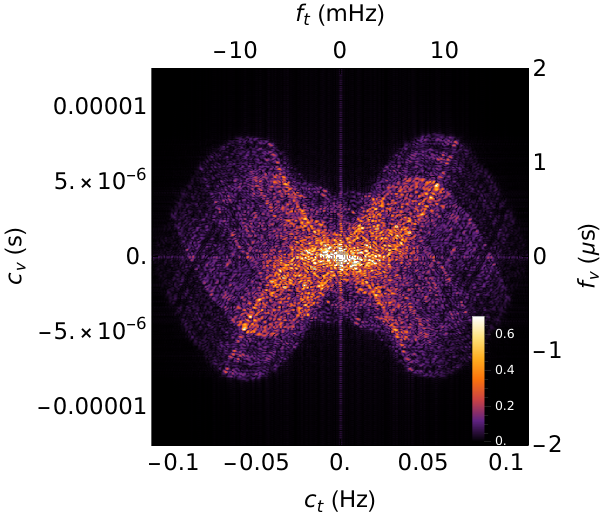}\\
\includegraphics[width=0.33\textwidth]{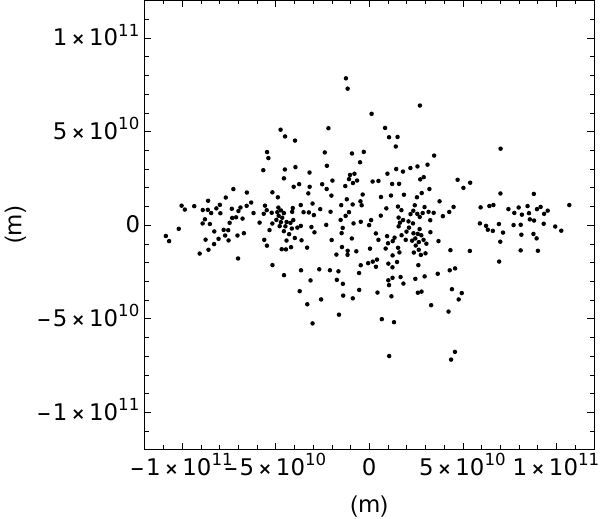}
\includegraphics[width=0.31\textwidth]{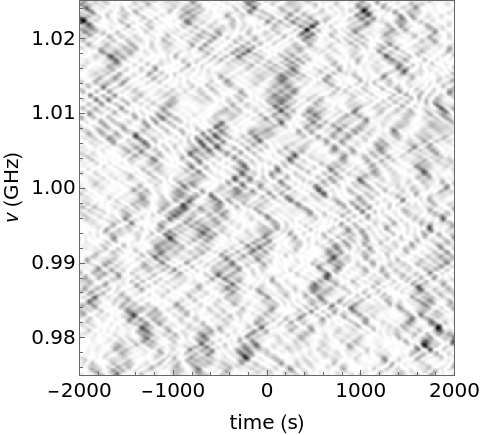}
\includegraphics[width=0.351\textwidth]{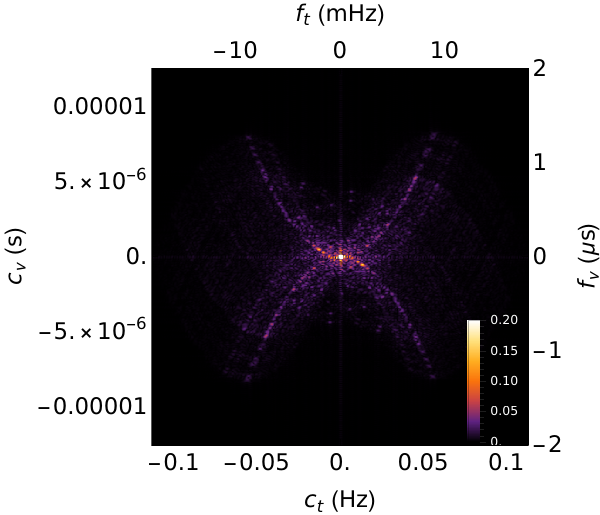}
\end{center}
\caption{\label{fig:eta}
Scatterer distribution in the $y-z$ plane (left column), dynamic spectra $H(t,\nu)$ and secondary spectra ($|\tilde{H}(c_t=2\pi f_t,c_\nu=2\pi f_\nu)|$ for two different values of $\beta$, corresponding to a change in contrast between main and inverted parabolic arcs.
Upper row $\beta=5\times 10^{18}$~m (for all scatterers), lower row $\beta=10^{18}$~m (for all scatterers), other parameters
$\nu_c=1$~GHz, $x_p=-1313$~pc, $x_o=788$~pc, $v_p=1000$~km/s. 
}
\end{figure}

\subsection{Main parabolic arc features}

The Fourier transform of the first term in Eq.~(\ref{eq:pspec}) describes interference between the direct path of the electric field from the pulsar to the observer and the path going through the ISM at a cloud centred at position $(0,y_i,z_i)$.
The Fourier integral comprises terms in the form 

\begin{equation}
\tilde{H}^{(i)}_1(c_t,c_\nu)=\beta_i
\int_{\nu-\Delta_\nu/2}^{\nu+\Delta_\nu/2}\int_{-\Delta_t/2}^{\Delta_t/2} 
\frac{\rme^{\rmi \nu c_\nu+\rmi t c_t} \left(\rme^{ \rmi k ( |\mathbf{r}_p-\mathbf{r}_i|
                     +|\mathbf{r}_i-\mathbf{r}_o|
                     -|\mathbf{r}_p-\mathbf{r}_o|)
                     }+\rme^{ -\rmi k ( |\mathbf{r}_p-\mathbf{r}_i|
                     +|\mathbf{r}_i-\mathbf{r}_o|
                     -|\mathbf{r}_p-\mathbf{r}_o|)
                     }\right)}
                     {|\mathbf{r}_p-\mathbf{r}_i| |\mathbf{r}_i-\mathbf{r}_o| |\mathbf{r}_p-\mathbf{r}_o|} \; \rmd\nu \; \rmd t.
\label{eq:interfer1}
\end{equation}
To evaluate the integrals in the last equation, we set $\mathbf{r}_p(t)=(x_p,v_p t,0)$, 
and $\mathbf{r}_o(t)=(x_o,\mathbf{v}_{o,y} t,\mathbf{v}_{o,z} t)$.
The expressions derived in this section include the possibility of arbitrary movements of pulsar, observer, and ISM.
The argument of the exponential functions is expanded around $x_p=-\infty$ and $x_o=\infty$ to first order. 
In addition, we consider only the first order in $t$ and $\nu$ around zero and ${\nu_c}$, respectively.
The denominator is taken to be constant.
Using the relation
\begin{equation}
    \int_{-\infty}^\infty \rmd x\; \rme^{i (k-k') x}=\frac{1}{2\pi}\delta(k-k').
\end{equation}
we obtain for the first exponential function
\begin{align}\label{eq:cti}
     &c_{t}^{(i)}  = \frac{2 \pi  y_i \nu _c \mathbf{v}_{p,y}}{c x_p}-\frac{2\pi y_i\nu_c\mathbf{v}_{o,y}}{cx_o}-\frac{2 \pi  \mathbf{v}_{o,z} {z_i} \nu _c}{c x_o}\\\label{eq:cnui}
     &c_{\nu}^{(i)}=-\frac{\pi  \left(x_o-x_p\right) \left(y_i^2+{z_i}^2\right)}{c x_o x_p}.
\end{align}
The second exponential function in Eq.~(\ref{eq:interfer1}) leads to the expressions in Eqs.~(\ref{eq:cti},\ref{eq:cnui}) with the replacement $c\rightarrow -c$.
These expressions agree with the ones in \cite{hill_pulsar_2003,hill_deflection_2005} and \cite {cordes_theory_2006}, which can be shown by introducing for $\pmb{\theta}$ and ${\mathbf{v}}_\perp$ the corresponding components.

These points lie on a parabolic arc, as seen by eliminating $y_i$ from the last equation and expressing $c_\nu$ as function of $c_t$ 
\begin{equation}\label{eq:mainpoints}
    c_\nu=\frac{\left(x_p-x_o\right)x_p}{\left(\mathbf{v}_{p,y}x_0-\mathbf{v}_{o,y}x_p\right)^2}\left[\sqrt{\frac{cx_o}{\pi}}\frac{ c_t}{2\nu _c}+\sqrt{\frac{\pi}{cx_o}}\mathbf{v}_{o,z}z_i\right]^2-\frac{\pi  z_i^2 \left(x_o-x_p\right)}{c x_o  x_p}
\end{equation}
For $z_i=0$ and $\mathbf{v}_{o,z}=0$ the last expression reduces to the parabolic arc expression crossing the origin of the $c_\nu,c_t$ coordinate system.
Eq.~(\ref{eq:mainpoints}) 
shows that $z_i$ induces a shift of the parabolic structures in the $c_\nu$-direction and $\mathbf{v}_{o,z}$ a shift in $c_t$-direction, moving the parabola away from the the origin of the $c_\nu,c_t$ coordinate system.
The corresponding expressions within the Fresnel-Kirchhoff approach (\cite{Walker2004}) for arbitrary two-dimensional scatterer positions and velocities  have been given in \cite[Fig.~A1]{xu_interstellar_2018} and in \cite{shi_morphology_2021}.
The $c_\nu$ coordinates of the points allow one to construct a projected spatial distribution of the scatterers along the $\mathbf{e}_y$ axis from the secondary spectra, or, for points clearly offset from the main parabola, to determine their $\mathbf{e}_z$ coordinate.
Note that \cite{cordes_theory_2006} introduce a $1/(2\pi)$-scaled variant of the conjugate quantities:
\begin{equation}
f_\nu=\frac{c_\nu}{2\pi},\quad f_t=\frac{c_t}{2\pi},    
\end{equation}
conventionally referred to as $\tau$ and $f_D$ in the literature, respectively.

A detailed quantitative description of the features requires to move beyond the linearised exponents and to evaluate the integrals in terms of special functions, see Appendix~\ref{app3}.
For simplicity of presentation,  we concentrate on the special case ${\bf{v}}_o=0$. 
For the main parabolic arc we obtain a trapezoid around the centre point $(c_{t}^{(i)}, c_{\nu}^{(i)})$ with vertices and magnitude
\begin{align}
    c_{t,\pm\mp}^{(i)} &= c_{t}^{(i)}
    \pm\frac{\pi  \Delta_\nu  y_i \mathbf{v}_{p,y}}{c x_p}
    \mp\frac{\pi  \Delta_\nu  x_o \mathbf{v}_{p,y}^2 \Delta _t}{2 c \left(x_o-x_p\right) x_p},
    \notag \\
    c_{\nu,\mp}^{(i)}  &= c_{\nu}^{(i)} \mp \frac{\pi \Delta_t y_i \mathbf{v}_{p,y}}{c x_p}+\frac{\pi  x_o \mathbf{v}_{p,y}^2 \Delta _t^2}{4 c x_o x_p-4 c x_p^2},\label{eq:trap}\\
    |\tilde{H}_1^{(i)}|    &= \left| \beta_i \frac{c}{\mathbf{v}_p x_o (x_o-x_p) y_i} \right|.\label{magni1}
\end{align}
Here, the $\pm$, $\mp$ in $c_{t,\pm\mp}^{(i)}$ refer to the four values of $c_{t}^{(i)}$ at the right/left and upper/lower boundaries of the trapezoid and the $\mp$ in $c_{\nu,\mp}^{(i)}$ to the upper/lower $c_{\nu}^{(i)}$-values at the borders.
Regions with similar electron density but further away from the line of sight will result in larger areas in the secondary spectrum with a magnitude proportional to the inverse distance $1/y_i$.
The extension of the Fourier window (and in general shape of the chosen window) changes the secondary spectra by affecting the size of the rectangular areas in the secondary spectra.
Each of these trapezoids comes with a complex phase leading to interference effects in the case of overlaps of different trapezoids.
The right panel of Fig.~\ref{fig:spectra} shows a close-up of the patterns with solid lines drawn according to Eq.~(\ref{eq:trap}).

\subsection{Inverted parabolic arcs}

The Fourier transform of the second term in Eq.~(\ref{eq:pspec}) arises from interference between two waves travelling through the ISM at positions $(y_i,z_i)$ and at $(y_j,z_j)$.
\begin{equation}
\tilde{H}_2^{(i,j)}(c_t,c_\nu)=\beta_i\beta_j
\int_{\nu-\Delta_\nu/2}^{\nu+\Delta_\nu/2}\int_{-\Delta_t/2}^{\Delta_t/2} 
\frac{\rme^{\rmi \nu c_\nu+\rmi t c_t} \rme^{ i k ( |\mathbf{r}_p-\mathbf{r}_i|
                     +|\mathbf{r}_i-\mathbf{r}_o|
                     -|\mathbf{r}_p-\mathbf{r}_j|
                     -|\mathbf{r}_j-\mathbf{r}_o| ) }}
                     {|\mathbf{r}_p-\mathbf{r}_i| |\mathbf{r}_i-\mathbf{r}_o| |\mathbf{r}_p-\mathbf{r}_j| |\mathbf{r}_j-\mathbf{r}_o|} \rmd\nu \rmd t
.\label{eq:interfer2}
\end{equation}
This expression is evaluated as in the last subsection and yields maxima at points
\begin{align}
    c_{t}^{(i,j)}  &=
    \frac{2 \pi  \nu _c \mathbf{v}_{p,y} \left(y_i-y_j\right)}{c x_p}-\frac{2\pi\nu_c \mathbf{v}_{o,y}\left(y_i-y_j\right)}{cx_o}
    +\frac{2 \pi  \nu _c \mathbf{v}_{o,z} \left(z_j-z_i\right)}{c x_o}\label{eq:invertedpoints0}
    \\\label{eq:invertedpoints}
    c_{\nu}^{(i,j)} &= -\frac{\pi  \left(x_o-x_p\right) \left(y_i^2-y_j^2+z_i^2-z_j^2\right)}{c x_o x_p}.
\end{align}
By the same replacements as described after Eq.\ (\ref{eq:cti}) the expressions Eqs.~(\ref{eq:invertedpoints0},\ref{eq:invertedpoints}) can be shown to be identical to the ones in \cite{hill_pulsar_2003,hill_deflection_2005,cordes_theory_2006}.
These points lie on inverted parabolic arcs and illuminate rectangular areas in the secondary spectra with vertices and magnitude given by 
\begin{align}
    c_{t,\pm}^{(i,j)}   &= c_t^{(i,j)}   \pm \frac{\pi \mathbf{v}_{p,y} (y_i-y_j)}{c x_p} \Delta_\nu, \notag \\ 
    c_{\nu,\mp}^{(i,j)} &= c_\nu^{(i,j)} \mp \frac{\pi \mathbf{v}_{p,y} (y_i-y_j)}{c x_p} \Delta_t, \label{eq:rect}\\
    |\tilde{H}_2^{(i,j)}| &= \beta_i \beta_j \left| \frac{c}{\mathbf{v}_{p,y} x_o^2 x_p (y_i-y_j)} \right|.\label{magni2}
\end{align}
The right panel of Fig.~\ref{fig:spectra} shows a close-up of the patterns with dashed rectangles drawn according to Eq.~(\ref{eq:rect}).
Thus, our analytic expressions (\ref{eq:trap},\ref{eq:rect}) are in excellent agreement with our numerics.
\begin{figure}
\begin{center}
\includegraphics[width=0.33\textwidth]{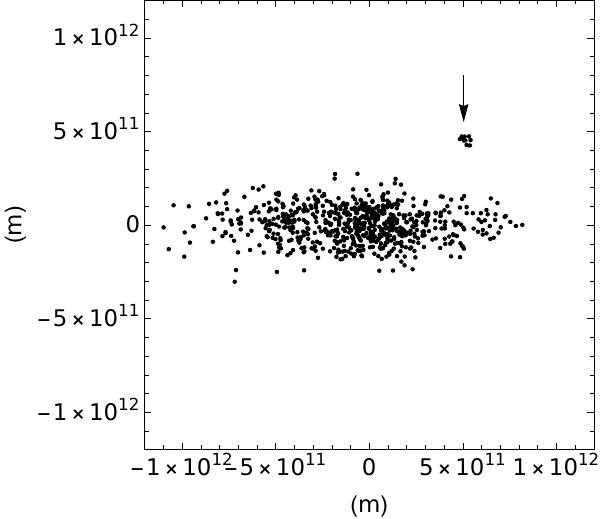}
\includegraphics[width=0.315\textwidth]{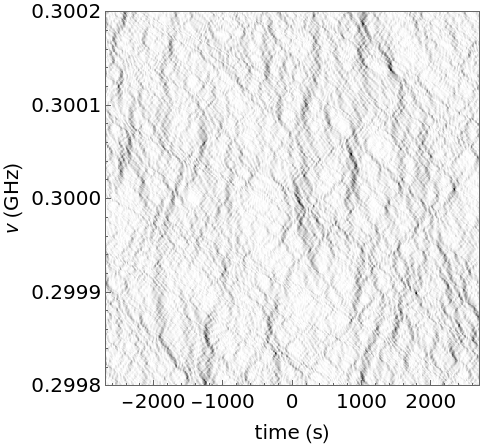}
\includegraphics[width=0.345\textwidth]{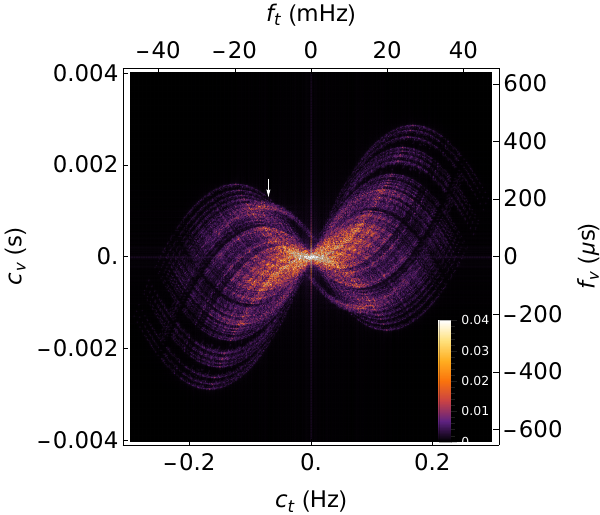}\\
\end{center}
\caption{\label{fig:brisken}
Left panel: example of an extended scatterer set with one separated scatterer region (black arrow), leading to the formation of an inverted arc offset from the main parabolic arc in the secondary spectrum $|\tilde{H}(c_t=2\pi f_t,c_\nu=2\pi f_\nu)|$ (right panel), marked by the white arrow. 
The dynamic spectrum $H(t,\nu)$ is shown in the centre panel.
Conjugate time $c_t=2\pi f_t$, conjugate frequency $c_\nu=2\pi f_\nu$. 
Parameters: $\nu_c=0.3$~GHz, $\beta=10^{18}$~m (for all scatterers), $x_p=-214$~pc, $x_o=429$~pc, $v_p=160$~km/s.}
\end{figure}
In the ``noodle model of scintillation arcs'' proposed by Gwinn a smearing of these structures \cite[Eqs. (62,63)]{gwinn_noodle_2019}  leads to partly similar expressions as our Eqs. (\ref{eq:cti}), (\ref{eq:rect}). 
The differences in the results originate from the differing approaches: whereas Gwinn analyses how in the expressions for the spot positions in (\ref{eq:cti},\ref{eq:cnui}) and (\ref{invertedpoints0},\ref{eq:invertedpoints}) change during the integrations in (\ref{eq:interfer1}) and (\ref{eq:interfer2}), respectively, we directly analyse the analytical results for (\ref{eq:interfer1}) and (\ref{eq:interfer2}).
Furthermore, the respective magnitudes (\ref{magni1},\ref{magni2}) and consequences with respect to the visibility of main vs. inverted arcs are not discussed in \cite{gwinn_noodle_2019}. 
The issue of resolution was described in \cite{Walker2004}. 
The authors obtain $\text{sinc}$-functions which lead to decreasing/increasing spot sizes in $c_t$ ($c_\nu$)-direction with increasing/decreasing $\Delta_t$ ($\Delta_\nu$). 
Existing techniques to combat this are performing the Fourier transform with respect to wavelength instead of frequency \cite{fallows_broadband_2014,reardon_precision_2020} or with respect to time times frequency instead of time \cite{sprenger__2020}.
We emphasise that our expressions derived in Appendix \ref{app3} describe both effects, the smearing and the resolution.

\subsection{Magnitudes of main and inverted parabolic arcs}

The ratio of the magnitude of an inverted parabolic arc structure compared to a main parabolic structure is given by the expression
\begin{equation}\label{eq:ratiointensity}
\frac{|\tilde{H}_2^{(i,j)}(c_t,c_\nu)|}{|\tilde{H}_1^{(i)}(c_t,c_\nu)|} = \beta_j \left|\frac{ y_i }{\left(y_i-y_j\right)}\frac{\left(x_o-x_p\right)}{x_o x_p}\right|.
\end{equation}
Eq.~(\ref{eq:ratiointensity}) implies an increased visibility of the inverted arc structures for higher plasma densities (corresponding to a larger value of $\beta$).
Some pulsars (see pulsar B1508+55 discussed by \cite{Sprenger2022} show a transient evolution of the secondary spectra at different epochs with less and more pronounced inverted arc structure.
For a pulsar located $1313$~pc from the ISM and observed at a distance of $788$~pc from the ISM, $|\frac{\left(x_o-x_p\right)}{x_o x_p}|=6.5\times 10^{-20}$~m$^{-1}$.
Setting $|\frac{ y_i }{\left(y_i-y_j\right)}|\approx 2$ gives a value of $\beta=7.5\times 10^{18}$~m to distribute equal intensities to the direct and inverted arc features.
This estimate is in general agreement with the transition from a single main arc ($\beta=10^{18}$~m) to inverted arcs ($\beta=5\times 10^{18}$~m) seen in Fig.~\ref{fig:eta}.
The Helmholtz equation describes the underlying physics in terms of a local change in the refractive index, and thus cannot distinguish between an increase or decrease in matter relative to an average background density.
If we take ionised gas (plasma) as the origin of the refractive index change, we can estimate the electron density of the ISM structures, since the $\beta$ parameter is directly related to the refractive index change.

For the model in Fig.\ \ref{fig:eta} we chose $N=317$ scattering clouds in order to avoid considerable overlap of the corresponding panels in the secondary spectrum.
For simplicity, all clouds are assigned the same $\beta$ value.
The parameter $\beta=1\times 10^{18}$~m could be realised with a Gaussian cloud with $a=5\times 10^8$~m and $n_{e,\text{peak}}=0.15$~cm$^{-3}$, which is about ten times the average electron density $n_0=0.015$~cm$^{-3}$  in our galaxy (\cite{ocker_electron_2020}).
In the large $N$-regime areas in the secondary spectra overlap and additional interference of the individual structures in the secondary spectrum occurs.
Refractive index changes might also be coming from neutral gas clouds.
The finite extension of all interference structures leads to further interference between overlapping rectangles and trapezoids as seen in the right panel of Fig.~\ref{fig:spectra}.
The interference causes smaller scale structures compared to the extension of the rectangular or trapezoidal areas.

Fig.~\ref{fig:brisken} displays the dynamic and secondary spectra of a ISM region with a split-off part (see black arrow in the left panel), causing an offset feature in the secondary spectra (white arrow in the right panel).
The split-off part produces shifted inverted arclets by including a clump of scatterers offset in the $z$-direction, but still in the same scattering screen (see Fig.~\ref{fig:brisken}, left panel).
Similar structures have been observed by \cite{brisken_100_2010} for pulsar B0834+06, 
but are attributed there to a lens-like concentration of plasma due to their different movement with wavelengths (see also \cite{Simard2018}), or to multiple screens (\cite{simard_disentangling_2019}, \cite{zhu_pulsar_2023}).

\section{Conclusion}

We show that the Born approximation to Green's function is a suitable method for computing dynamic and secondary spectra of pulsar signals.
The theoretical description does not use the Fresnel-Kirchhoff diffraction integral.
Our method paves the way for the effective extraction of physical parameters such as the refractive index change and the spatial structure of the ISM from secondary spectra.
The main and inverted arcs seen in secondary spectra are obtained without assuming a quasi one-dimensional structure of the ISM. 
Furthermore, within this approach, we are able to compute and analyse the spectra with high precision numerically, as well as explain them analytically.
The method can be generalised in several directions, e.g., it is straightforward to describe configurations containing multiple screens or screens extended in three dimensions.  
The Green's function method could also be applied to plasma lens structures, as e.g.\ the Gaussian plasma lens (\cite{clegg_gaussian_1998}), not considered here.

\section*{Acknowledgements}

This work benefited from the NVIDIA Academic Hardware Grant \textit{Simulating branched flow with tensor processing units}, PI Kramer (2022).

\section*{Data availability}

The data underlying this article will be shared on reasonable request to the corresponding author.


\appendix

\section{Relation to Kirchhoff-Fresnel theory}\label{app1}

Pulsar scintillations have been discussed previously using Kirchhoff-Fresnel theory, see \cite[chapter 8.3]{born_principles_2019}.
In addition a phase-changing screen is introduced around the origin with screen coordinates $(0,y',z')$, leading to the expression \cite[Eq.~(2.1)]{narayan_physics_1992}:
\begin{equation}\label{eq:KDI}
    E(\mathbf{r}_o)=k \frac{\rme^{-\rmi \pi/2}}{2\pi D}
    \iint \exp\left[\rmi \varphi(y',z')+\rmi k \frac{{(y'-y_o)}^2+{(z'-z_o)}^2}{2 D}\right] \rmd y' \rmd z', \quad D=|\mathbf{r_o}|.
\end{equation}
For a model based on stripes of phase changing scatterers perpendicular to the line-of-sight between pulsar and observer \cite{gwinn_noodle_2019} provides a detailed analysis based on Kirchhoff-Fresnel integrals.
In contrast to our expression in Eq.~(\ref{eq:greenE}), Kirchhoff-Fresnel theory does not contain the contributions from the scattering at the ISM explicitly as scattering volumes ($\beta_i$ in Eq.~(\ref{eq:eta})) and thereby does not distinguish the contributions of the different terms in Eqs.~(\ref{eq:H}), (\ref{eq:pspec}) which include the unobstruced path.
To recover the unobstructed path, \cite{gwinn_noodle_2019} introduces an additional ``no screen'' term, relative to which any phase changes are considered.
Besides a stripe model, also various plasma lenses are discussed by \cite{jow_regimes_2023} in the context of Kirchhoff-Fresnel theory and a further perturbative expansion of the integrand in Eq.~(\ref{eq:KDI}) is given.
A treatment of a Gaussian lens is shown in \cite[Fig.~1c]{aidala_imaging_2007} and \cite[Fig.~2]{jow_regimes_2023}.
In contrast to the Born approximation with a real-valued scattering potential across a volume, the starting point for Kirchhoff-Fresnel theory is a phase change of the electric field caused by refractive index changes projected on a plane.
Whereas within the Born approximation considered here, the scattering of radiation at the ISM causes spherical waves originating from all scatterers in three dimensions, in the approximation of Fresnel Kirchhoff spherical waves originate only from a two dimensional plane.

\section{Dimensionality of the ISM}\label{app2}

In Fig.~\ref{fig:pulsarscheme} we depicted the ISM as a 3-dimensional cloud in space. However, throughout the paper it was always considered to be 2-dimensional. To justify this restriction to 2 dimensions, we show here that the effect of an additional extension along the $x$-axis is negligible: 

Therefore we compute the length difference $\Delta s$ between the path along the line-of-sight starting at the pulsar located at $(x_p,0,0)$ and ending at the observer at $(x_o,0,0)$ and another path starting at the pulsar, going to the ISM offset from the line-of-sight at  $(0,y_0,0)$ and from there to the observer is given by    
\begin{equation}\label{eq1}
       \Delta s =\sqrt{x_o^2+y_0^2}+\sqrt{x_p^2+y_0^2}-|x_o|-|x_p|\approx\frac{y_0^2}{2}\left(\frac{1}{|x_o|}+\frac{1}{|x_p|}\right),
\end{equation}
where the last relation holds in the limit of $|x_0|\gg y_0$ and $|x_p|\gg y_0$. Shifting the scatterer along the line-of-sight from $(0,y_0,0)$ to $(\Delta x,y_0,0)$ changes the result in Eq.\ (\ref{eq1}) to  
\begin{equation}\label{eq4}
       \Delta s \approx\frac{y_0^2}{2}\left(\frac{1}{|x_o|+\Delta x}+\frac{1}{|x_p|-\Delta x}\right)\approx\frac{y_0^2}{2}\left(\frac{1}{|x_o|}+\frac{1}{|x_p|}-\frac{\Delta x}{|x_o|^2}+\frac{\Delta x}{|x_p|^2}\right).
\end{equation}
As long as $|\Delta x|\ll|x_o|$ and $|\Delta x|\ll|x_p|$, the impact of $\Delta x$ on the interference is quite small, i.e.\ for the setup in Fig.~\ref{fig:eta} and a scatterer at a distance of $1$ a.u.\ away from the line of sight, a 170,000 a.u.\ shift along $x$ gives a $0.3$~m change in path difference (radio wavelength $\lambda=0.3$~m at 1~GHz).
In the case $|x_o|=|x_p|$, there is no first order dependence on $\Delta x$. 
To produce a change of $\Delta s$ of 1~meter at $|x_o|=500$~pc, $\Delta x$ can extend up to $13$~pc, corresponding to 3 percent of the LOS.

Due to these observations we do not consider explicitly the extension of the ISM in $x$-direction. 
It would be straightforward to include this effect in our calculation and extend our result to that case.

In contrast to a change of the $x$ position of the scatterer, a change in $y$ direction by $\Delta y$ leads to 
\begin{equation}\label{eq5}
       \Delta s \approx\frac{(y_0+\Delta y)^2}{2}\left(\frac{1}{|x_o|}+\frac{1}{|x_p|}\right) \approx \frac{y_0^2+2y_0\Delta y}{2}\left(\frac{1}{|x_o|}+\frac{1}{|x_p|}\right).
\end{equation}
Comparing the effects of $\Delta x$ in Eq.~(\ref{eq4}) and of $\Delta y$ in Eq.~(\ref{eq5}), respectively, we see that under the assumption $|x_o|\approx|x_p|$ the effect of $\Delta x$ is by a factor $y_0/|x_o|$ smaller than the effect of $\Delta y$.

\section{Born approximation for a Gaussian distribution of electrons}\label{app:Born}

Consider the electron density of the $i$th cloud
\begin{equation}
n_{e,i}(\mathbf{r''})=n_{e,\text{peak},i}  \exp{\left(-\frac{{(\mathbf{r}''-\mathbf{r_i})}^2}{2 a^2}\right)},
\end{equation}
To evaluate the integral of the first order Born approximation
\begin{equation}\label{eq:GB}
G_\text{Born}(\mathbf{r},\mathbf{r}')=
\frac{\rme^{\rmi k |\mathbf{r}-\mathbf{r}'|}}{|\mathbf{r}-\mathbf{r}'|}
-\frac{e^2}{4\pi c^2 \epsilon_0 m_e }
\int_{\text{cloud}_i} \rmd\mathbf{r}''\; 
\frac{\rme^{\rmi k |\mathbf{r}-\mathbf{r}''|}}{|\mathbf{r}''-\mathbf{r}'|}
n_{e,i}(\mathbf{r}'')
\frac{\rme^{\rmi k |\mathbf{r}-\mathbf{r}''|}}{|\mathbf{r}''-\mathbf{r}'|},
\end{equation}
we use the propagator representation of the free Green's function, which has a Gaussian kernel:
\begin{equation}
\frac{\rme^{\rmi k |\mathbf{r}-\mathbf{r}''|}}{|\mathbf{r}-\mathbf{r}''|}=\frac{4\pi}{\rmi}\int_0^\infty \rmd t\;
 {\left(\frac{1}{4\pi \rmi t}\right)}^{3/2}\exp\left(\rmi\frac{{(\mathbf{r}-\mathbf{r}'')}^2}{4t}\right)\rme^{\rmi k^2 t}
\end{equation}
For definiteness we set $\mathbf{r}_i=({0,y_i,0})$, $\mathbf{r}=(x,0,0)$, and $\mathbf{r}'=(x',0,0)$.
This allows us to perform the spatial integration $\mathbf{r}''$ analytically and the remaining integral reads
\begin{equation}
- 2 \rmi \sqrt{\pi } a^3 n_{e,\text{peak},i}  \int_0^\infty \rmd t' \int_0^\infty \rmd t'' \rme^{\rmi k^2 t'} \rme^{\rmi k^2 t''}
\frac{\exp \left(\frac{a^2 (x-x')^2+2
   \rmi \left(t' \left(x'^2+y_i^2\right)+t''
   \left(x^2+y_i^2\right)\right)}{8 t' t''-4 \rmi a^2
   (t'+t'')}\right)}{\left(2 t' t''-\rmi a^2
   (t'+t'')\right)^{3/2}}.
\end{equation}
Expanding the integrand in a power series around $a=0$ and integrating term by term yields
\begin{equation}
 G_\text{Born}(\mathbf{r},\mathbf{r}')=
 \frac{\rme^{\rmi k |\mathbf{r}-\mathbf{r}'|}}{|\mathbf{r}-\mathbf{r}'|}
 +\beta_i\frac{e^{\rmi k \left(\sqrt{\xi^2}+\sqrt{\xi'^2}\right)}}{\sqrt{\xi^2
   \xi'^2}}\label{eq:bornlast}
   -\beta_i\frac{a^2 e^{\rmi k (\xi+\xi')} \left(k^2 \xi^2 \xi'^2+x
   x' (k \xi+\rmi) (k \xi'+\rmi)+y_i^2 (k \xi+\rmi) (k
   \xi'+\rmi)\right)}{\xi^3 \xi'^3}+\ldots,
\end{equation}
where we introduced $\xi^2=x^2+y_i^2$, $\xi'^2=x'^2+y_i^2$ and used the definition of $\beta_i$ (Eq.~\ref{eq:eta}).
The first term is the direct path from the pulsar to the observer,
the second term is identical to the interaction of the pulsar pulse with a point scattering source obtained by contracting the Gaussian cloud, and the third term leads to a direction dependent scattering amplitude.
We conclude that the contraction of the Gaussian cloud to a point is a valid approximation if the last term in Eq.~(\ref{eq:bornlast}) can be neglected; otherwise, it should be included and leads to a diminishing effect of scattering clouds off the line of sight.

\section{Saddle point evaluation}\label{app3}

For deriving the finite extensions of the interference regions in the secondary spectra, it is convenient to introduce the effective perpendicular velocity of the interstellar medium
\begin{equation}
    v^\text{eff}=-\mathbf{v}_p \frac{x_o}{x_o-x_p}.
\end{equation}
In this coordinate frame the pulsar and observer are kept at rest.

\subsection{Main parabolic arc}

The saddle points of the integral reveal the trapezoidal area in the secondary spectra. 
We start from Eq.\ (\ref{eq:interfer1}) and expand the arguments of the exponential functions to the first order around $x_p=-\infty$ and $x_o=\infty$, which yields 
\begin{align}
\bar{H}^{(i)}_1(c_t,c_\nu)=\beta_i
\int_{\nu-\Delta_\nu/2}^{\nu+\Delta_\nu/2}\int_{-\Delta_t/2}^{\Delta_t/2} 
\frac{\rme^{\rmi \nu c_\nu+\rmi t c_t} \rme^{ \rmi \pi\nu ( (y_i+v^{\text{eff}}t)^2(1/x_p-1/x_o)/c
                     }}
                     {(x_o-x_p)x_ox_p} \; \rmd\nu \; \rmd t.
\end{align}
We perform one of the integrals analytically, while we expand the integrand of the remaining Fourier transform using $\text{erfi}(z)=\frac{2}{\rmi\sqrt{\pi}}\int_0^{\rmi z}\rme^{-t^2}\rmd t\approx -\rmi + \frac{\rme^{z^2}}{\sqrt{\pi}z}$. 
%
In this context we restrict to the first summand in Eq.~(\ref{eq:interfer1}) and denote the corresponding contribution by a bar instead of a tilde, the contribution from the second summand is obtained by the replacement $c\to-c$ from the first one
\begin{align}\label{eq:secspecanalyticmain}
    \bar{H}_1^{(i)}(c_t,c_\nu)&=-\int_{\nu_c-\Delta_\nu/2}^{\nu_c+\Delta_\nu/2}\frac{(-1)^{3/4} \beta_i \exp \left(-\frac{1}{4} \rmi \left(-4 \nu  c_{\nu }+\frac{c x_o c_t^2 x_p}{\pi  \nu  x_o
   {(v^\text{eff})}^2-\pi  \nu  {(v^\text{eff})}^2 x_p}+\frac{4 y_i c_t}{{v^\text{eff}}}\right)\right)}{2 {v^\text{eff}} \left(x_o-x_p\right){}^{3/2}
   \sqrt{\frac{\nu  x_o x_p}{c}}}\\
    &\quad\quad\bigg[\text{erfi}\left(\frac{\sqrt[4]{-1} \left(c x_o c_t x_p+\pi  \nu  {v^\text{eff}} \left(x_o-x_p\right) \left({v^\text{eff}} \Delta _t+2
   y_i\right)\right)}{2 \sqrt{\pi } {v^\text{eff}} \sqrt{c \nu  x_o \left(x_o-x_p\right)
   x_p}}\right)\nonumber\\
   &\quad\quad-\text{erfi}\left(\frac{\sqrt[4]{-1} \left(c x_o c_t x_p+\pi  \nu  {v^\text{eff}} \left(x_o-x_p\right) \left(2
   y_i-{v^\text{eff}} \Delta _t\right)\right)}{2 \sqrt{\pi } {v^\text{eff}} \sqrt{c \nu  x_o \left(x_o-x_p\right) x_p}}\right)\bigg]\rmd \nu \nonumber\\
   &\approx
   \int_{\nu_c-\Delta_\nu/2}^{\nu_c+\Delta_\nu/2} \bigg[-\frac{\rmi c \beta_i  \exp \left(\frac{1}{4} \rmi \left(4 \nu  c_{\nu }+\frac{\pi  \nu  \left(x_o-x_p\right) \left({v^\text{eff}} \Delta _t-2
   y_i\right){}^2}{c x_o x_p}-2 c_t \Delta _t\right)\right)}{\left(x_o-x_p\right) \left(c x_o c_t x_p+\pi  \nu  {v^\text{eff}} \left(x_o-x_p\right) \left(2 y_i-{v^\text{eff}} \Delta
   _t\right)\right)}
   \\
   &
   +\frac{i c \beta_i  \exp \left(\frac{1}{4} \rmi \left(4 \nu  c_{\nu }+\frac{\pi  \nu  \left(x_o-x_p\right) \left({v^\text{eff}}
   \Delta _t+2 y_i\right){}^2}{c x_o x_p}+2 c_t \Delta _t\right)\right)}{\left(x_o-x_p\right) \left(c x_o c_t x_p+\pi 
   \nu  {v^\text{eff}} \left(x_o-x_p\right) \left({v^\text{eff}} \Delta _t+2 y_i\right)\right)}\bigg] \rmd \nu \nonumber.
\end{align}
The borders of the trapezoid along the $c_\nu$-axis are determined by the condition that the first derivative of the argument of the exponential functions with respect to $\nu$ becomes zero for $-\Delta_t/2\leq t\leq \Delta_t/2$, i.e.\ that the stationary point of the $\nu$ integral lies for $-\Delta_t/2\leq t\leq \Delta_t/2$ in the integration domain.
For the other edges of the trapezoid we first perform the integration over the frequency domain and determine the time $t_{sp}$ when the first derivative of the exponential function with respect to time vanishes. 
The requirement that the saddle point occurs in the interval  $-\Delta_t/2 \le t_{sp} \le \Delta_t/2$
determines the edges of the trapezoid:
\begin{align}
c_{t,\pm\pm\mp}^{(i)}&=\pm\frac{\pi  {v^\text{eff}} \left(x_o-x_p\right) \left(\nu_c\pm\frac{\Delta \nu }{2}\right) \left({v^\text{eff}} \Delta _t \mp 2
   y_i\right)}{c x_o x_p}\\
c_{t,\mp\pm\pm}^{(i)}&=\mp\frac{\pi  {v^\text{eff}} \left(x_o-x_p\right) \left(\nu_c\pm\frac{\Delta \nu }{2}\right) \left({v^\text{eff}} \Delta _t \pm 2
   y_i\right)}{c x_o x_p}\\
c_{\nu,\mp}^{(i)}&=-\frac{\pi  \left(x_o-x_p\right) \left({v^\text{eff}} \Delta_t \mp 2
   y_i\right){}^2}{4 c x_o x_p}.
\end{align}
The absolute value within the trapezoid is approximated by evaluating the magnitude at the central point using the residue theorem:
\begin{equation}
    |\bar{H}_1^{(i)}(c_t,c_\nu)|=\left|\frac{c\;\beta_i }{y_i {v^\text{eff}} \left(x_o-x_p\right){}^2}\right|.
\end{equation}

\subsection{Inverted parabolic arcs}

The rectangular area of the interference pattern is determined by analytically evaluating Eq.~(\ref{eq:interfer2}) in terms of exponential integral functions $\text{Ei}(z)=\int_{-z}^\infty e^{-t}/t \; \rmd t$:
\begin{align}
\tilde{H}_2^{(i,j)}(c_t,c_\nu)
&=\chi\bigg[
\text{Ei}\left(
\frac{\rmi \left(c_t-c_{t,+}\right) \left(c_{t,+} \left(-{v^\text{eff}} \Delta _t+y_i+y_j\right)+c_{\nu } {v^\text{eff}} \left(\Delta \nu -2 \nu _c\right)\right)}{2 {v^\text{eff}} c_{t,+}}
\right)\notag\\\notag
&\,\;-\text{Ei}\left(
\frac{\rmi \left(c_t-c_{t,+}\right) \left(c_{t,+} \left(+{v^\text{eff}} \Delta _t+y_i+y_j\right)+c_{\nu } {v^\text{eff}} \left(\Delta \nu -2 \nu _c\right)\right)}{2 {v^\text{eff}} c_{t,+}}
\right)\\\notag
&\,\;-\text{Ei}\left(
\frac{\rmi \left(c_{\nu }-c_{\nu,-}\right) \left({v^\text{eff}} \left(2 \nu _c+\Delta \nu \right) c_{\nu,-}-c_t \left(-{v^\text{eff}} \Delta _t+y_i+y_j\right)\right)}{2 {v^\text{eff}} c_{\nu,-}}
\right)\\\label{eq:ei}
&\,\;+\text{Ei}\left(
\frac{\rmi \left(c_{\nu }-c_{\nu,+}\right) \left({v^\text{eff}} \left(2 \nu _c+\Delta \nu \right) c_{\nu,+}-c_t \left(+{v^\text{eff}} \Delta _t+y_i+y_j\right)\right)}{2 {v^\text{eff}} c_{\nu,+}}
\right)
-{4\rmi\pi} {C(c_t,c_\nu)}
\bigg],
\end{align}
where $C$ is the characteristic function assuming the value 1 inside the rectangular area and 0 outside. 
In addition
\begin{equation}
\chi=\beta_i\beta_j \frac{\rmi c \exp \left(\frac{\rmi c x_o c_{\nu } c_t x_p}{2 \pi  \left(y_i-y_j\right) {v^\text{eff}} \left(x_o-x_p\right)}-\frac{\rmi \left(y_i+y_j\right) c_t}{2 {v^\text{eff}}}\right)}{2 \pi  \left(y_i-y_j\right) {v^\text{eff}} \left(x_o-x_p\right) x_o x_p}.
\end{equation}
The vertices of the rectangle are determined by the pole of the exponential integral function when the argument approaches zero. 
The graph of the exponential integral changes substantially when approaching the singularity \cite[\S~6.3]{olver_nist_2010}, leading to a sharp change in the function $\tilde{H}_2^{(i,j)}(c_t,c_\nu)$.
This condition yields a rectangular area with extensions
\begin{align}
c_{t,\pm}^{(i,j)}&=\frac{\pi  \left(y_i-y_j\right) {v^\text{eff}} \left(x_o-x_p\right) \left(2 \nu _c \pm \Delta \nu\right)}{c x_o x_p}\\
c_{\nu,\pm}^{(i,j)}&=\frac{\pi  \left(y_i-y_j\right) \left(x_o-x_p\right) \left(\pm{v^\text{eff}} \Delta _t+y_i+y_j\right)}{c x_o x_p}.
\end{align}
The absolute value in this area is determined by evaluating Eq.~(\ref{eq:ei}) at the centre point:
\begin{equation}
    |\tilde{H}_2^{(i,j)}(c_t,c_\nu)|=\beta_i\beta_j c \frac{2\,\text{Si} 
    \left(\frac{\pi {v^\text{eff}}\left(x_o-x_p\right) \left(y_i-y_j\right) \Delta t \Delta \nu}{2 c x_o x_p}\right)}{\pi {v^\text{eff}} x_o x_p (x_o-x_p) (y_i-y_j)}
    \approx 
    \frac{\beta_i\beta_j c}{\left(y_i-y_j\right) {v^\text{eff}} \left(x_o-x_p\right) x_o x_p},
\end{equation}
with 
$\text{Si}(z)=\int_0^z \frac{\sin(t)}{t} \rmd t$.

\section{Limit in terms of $\text{sinc}$-functions}\label{app4}

In this appendix we show that the features in the secondary spectra obtained for an expansion with respect to large distances of pulsar and observer (see App.~\ref{app3}) differ from the first order expansion in time and frequency described by \cite{Walker2004}, Eq.~(11).

\begin{figure}
    \centering
    \includegraphics[width=0.4\textwidth]{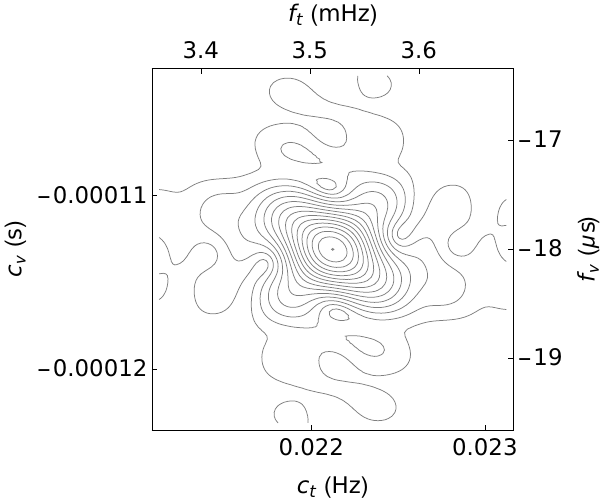}
    \includegraphics[width=0.4\textwidth]{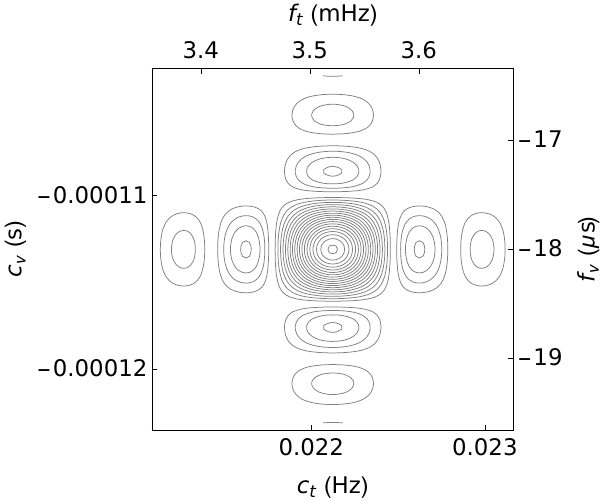}
    \caption{Left panel: Secondary spectrum $|\tilde{H}(c_t=2\pi f_t,c_\nu=2\pi f_\nu)|$ according to Eq.~(\ref{eq:ei}), right panel: linear approximation of the spectrum using Eq.~(\ref{eq:sinc}). 
    The numerical evaluation using Eq.~(\ref{eq:H1}) agrees with the left panel.
    Same parameters as in Fig.~\ref{fig:spectra}.
    }
    \label{fig:pillow20}
\end{figure}

 To simplify the derivation of the first order approximation with respect to time and frequency, we consider $z_i=0$, ${\mathbf v}_{o,z}=0$ and show here the corresponding derivation for the main parabolic arcs, the corresponding one for the inverted arcs follows similar steps. 
 We start from Eq.\ (\ref{eq:interfer1}) and again expand the arguments of the exponential functions to the first order around $x_p=-\infty$ and $x_o=\infty$, which yields 
\begin{align}
\bar{H}^{(i)}_1(c_t,c_\nu)=\beta_i
\int_{\nu-\Delta_\nu/2}^{\nu+\Delta_\nu/2}\int_{-\Delta_t/2}^{\Delta_t/2} 
\frac{\rme^{\rmi \nu c_\nu+\rmi t c_t} \rme^{ \rmi 2\pi\nu ( (y_i-\mathbf{v}_pt)^2/(2|x_p|)+y_i^2/(2x_o)-(\mathbf{v}_pt)^2/(2(x_o-x_p)))/c
                     }}
                     {-(x_o-x_p)x_ox_p} \; \rmd\nu \; \rmd t.
\end{align}
The integral with respect to $\nu$ is calculated first, giving
 \begin{align}
\bar{H}^{(i)}_1(c_t,c_\nu)=&\beta_i
\int_{-\Delta_t/2}^{\Delta_t/2} 
\frac{\rme^{\rmi \nu c_\nu+\rmi t c_t} \rme^{ \rmi 2\pi\nu ( (y_i-\mathbf{v}_pt)^2/(2|x_p|)+y_i^2/(2x_o)-(\mathbf{v}_pt)^2/(2(x_o-x_p)))/c
                     }}
                     {-(x_o-x_p)x_ox_p}\nonumber\\ &\frac{\sin \left( c_\nu\Delta_\nu/2+ \pi\Delta_\nu ( (y_i-\mathbf{v}_pt)^2/(2|x_p|)+y_i^2/(2x_o)-(\mathbf{v}_pt)^2/(2(x_o-x_p)))/c
                     \right)}{c_\nu/2+ \pi ( (y_i-\mathbf{v}_pt)^2/(2|x_p|)+y_i^2/(2x_o)-(\mathbf{v}_pt)^2/(2(x_o-x_p)))/c} \;  \; \rmd t.
\end{align}
To arrive at the result in \cite[Eq.~(11)]{Walker2004}, before performing the final $t$ integral, approximations must be made: the argument of the exponential in $t$ is linearised, in the second factor of the last equation the integration variable $t$ is replaced by its mean zero. The $t$ integral is then performed, giving
\begin{equation}\label{eq:sinc}
\bar{H}^{(i)}_1(c_t,c_\nu)=\beta_i
 \frac{\rme^{\rmi \nu c_\nu} \rme^{ \rmi 2\pi\nu ( y_i^2/(2|x_p|)+y_i^2/(2x_o)/c
                     }}
                     {-(x_o-x_p)x_ox_p}\Delta_t \Delta_\nu \operatorname{sinc}\left(\frac{c_t\Delta_t}{2}-\pi\frac{\nu y_i\mathbf{v}_p\Delta_t}{|x_p|c}\right)
\operatorname{sinc} \left( \frac{c_\nu\Delta_\nu}{2}+ \pi \frac{\Delta_\nu y_i^2}{c} \bigg( \frac{1}{2|x_p|}+\frac{1}{2x_o} \bigg)\right),
\end{equation}
which is the desired result in terms of $\text{sinc}$ functions, see \cite{Walker2004}. 
The resulting contour plot of a single scattering region using either the large distances expansions, Eq.~(\ref{eq:ei}), or the linearised time and frequency expression, Eq.~(\ref{eq:sinc}), are compared in Fig.~\ref{fig:pillow20}.
The numerical evaluation of Eq.~(\ref{eq:H1}) agrees with the large distance expansion, Eq.~(\ref{eq:ei}).

\section{Linear vs.\ log scale}

\begin{figure}
    \centering
    \includegraphics[width=0.45\textwidth]{figures/fig_pillows_all_Manuscript_FinalFig_3A_1__2024-05-01T10_30_00.pdf}
    \hfill\includegraphics[width=0.45\textwidth]{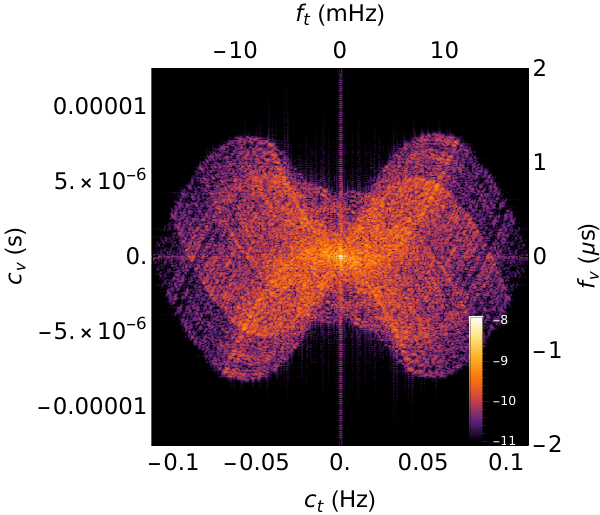}\\
    \includegraphics[width=0.45\textwidth]{figures/fig_pillows_all_Manuscript_FinalFig_3B_1__2024-05-01T10_20_25.pdf}
    \hfill\includegraphics[width=0.45\textwidth]{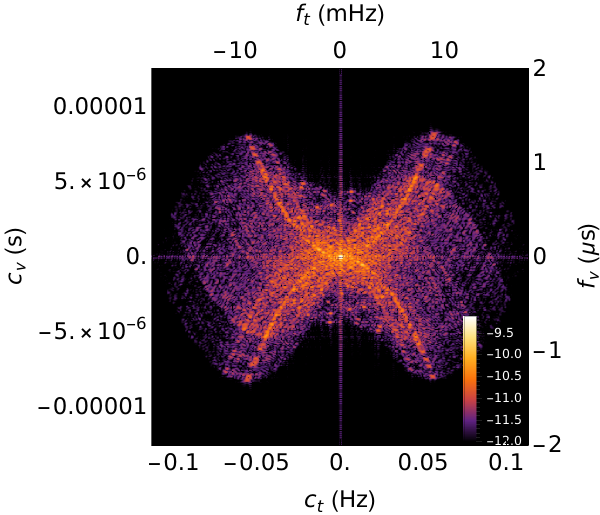}\\
    \includegraphics[width=0.47\textwidth]{figures/fig_pillows_all_Manuscript_FinalFig_4_1__2024-05-01T09_00_11.pdf}
    \hfill\includegraphics[width=0.47\textwidth]{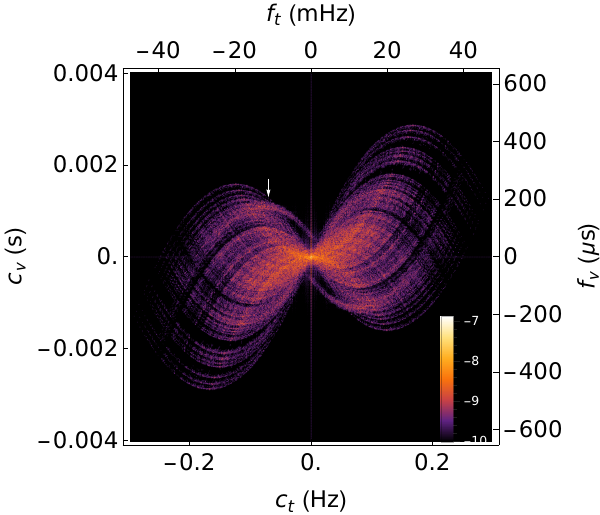}
    \caption{Left panels: Secondary spectra $|\tilde{H}(c_t=2\pi f_t,c_\nu=2\pi f_\nu)|$ shown in Figs.~\ref{fig:eta},\ref{fig:brisken} on a linear scale, right panels: logarithmic scale of the same data after subtracting the mean of each row.}
    \label{fig:logscale}
\end{figure}

Fig.~\ref{fig:logscale} shows the secondary spectra on linear and logarithmic intensity scales.
The logarithmic scale additionally removes the mean value of the intensity in each row and is commonly used to display secondary spectra derived from observations \cite{Stinebring2022}. In contrast to that, here we use the linear scale in order to directly compare our results with the analytical predictions. 

\end{document}